\documentclass[11pt]{article} 
\usepackage{rldmsubmit,palatino}
\usepackage{graphicx}
\usepackage{amsmath,amssymb}
\usepackage{listings}

\usepackage{algorithm}
\usepackage{algorithmic}
\usepackage{tikz}
\usetikzlibrary{arrows,positioning,shapes,calc}

\title{Why am I seeing this? Towards recognizing social media recommender systems with missing recommendations}

\author{
Sabrina Guidotti, Sabrina Patania, Giuseppe Vizzari, Dimitri Ognibene\\
Department of Psychology\\
University of Milan-Bicocca\\
Milan, Italy\\
\texttt{dimitri.ognibene@unimib.it} \\
\And
Gregor Donabauer, Udo Kruschwitz \\
Information Science \\
University of Regensburg \\
Regensburg, Germany \\
\And
Davide Taibi \\
ITD\\
CNR \\
Palermo, Italy \\
}

\begin{document}

\maketitle

\begin{abstract}

Social media plays a crucial role in shaping society, often amplifying polarization and facilitating the spread of misinformation. These effects arise from complex dynamics involving user interactions, individual characteristics, and recommender algorithms that drive content selection. Recommender systems, which significantly shape the content users see and decisions they make, present an opportunity for intervention and regulation. However, measuring their impact is challenging due to algorithmic opacity and the limited availability of data on their outputs. To effectively model social media users’ decision-making processes, it is crucial to recognize the recommender system adopted by the platform.

This work introduces a method for \textit{Automatic Recommender Recognition} using Graph Neural Networks (GNNs), relying exclusively on network structure and observed user behavior. To infer the \textit{hidden} recommender system, we first train a \textit{Recommender Neutral User model} (RNU) using a GNN and an adapted hindsight academic network recommender, hypothesized to reduce reliance on the actual recommender underlying the data. We then create several \textit{Recommender Hypothesis-specific Synthetic Datasets} $(RHSD(r))$ by combining the RNU with different known recommender systems, generating new ground truths for testing. Finally, we train \textit{Recommender Hypothesis-specific User models} $(RHU(r{\prime}))$ under varying hypotheses for the underlying recommender system $r{\prime}$, comparing the candidate recommender $r{\prime}$ with the original $r$ used to generate $RHSD(r)$.

Our approach allows for accurate detection of hidden recommenders and assessment of their influence on user behavior. Unlike audit methodologies, it directly captures actual system behavior without requiring ad hoc experiments, which often fail to reflect real social media platforms. By addressing how recommenders shape user behavior, this study provides critical insights into their societal impact, aiding efforts to mitigate polarization and misinformation.

\end{abstract}

\acknowledgements{We acknowledge support from the Volkswagen Foundation for the project \textit{Developing an Artificial Social Childhood (ASC) to improve AI causal reasoning, information gathering and decision making}, Ref.: 9E530.}


\newpage
\section{Introduction}

The ethical and societal implications of social media recommender systems have understandably drawn considerable attention. Research suggests that algorithms focused on engagement can inadvertently spread misinformation~\cite{tommasel2022recommender, bagchi2024social} and foster phenomena like filter bubbles echo chambers~\cite{almourad2020defining,gillani2018me,nikolov2015measuring} or backfire and extensive online firestorms~\cite{pfeffer2014understanding}. Filter bubbles limit users' exposure to diverse perspectives by favoring content that aligns with their beliefs~\cite{nikolov2015measuring}, while echo chambers intensify this effect by reinforcing shared ideas~\cite{gillani2018me,echo_chambers}. These trends highlight the need for transparency and accountability in recommender algorithms~\cite{ognibene2023challenging,ognibene2023moving}.

Various auditing techniques have been developed to examine these systems' biases and societal impacts. White-box auditing necessitates access and often transparency and interpretability of recommenders, requirements hardly met by current deep learning based recommenders. Black-box auditing, which analyzes user interactions without internal system access, has revealed biases that can amplify polarizing content~\cite{ribeiro2020auditing}, particularly on platforms like YouTube~\cite{tufekci2018youtube}. However, it often relies on ad-hoc tests and interventions that may not reflect the platform dynamic at full scale. Fairness-focused audits have also shown that recommendation systems may disproportionately impact certain groups, often perpetuating societal biases~\cite{sapiezynski2019colorblind}. These insights have fueled calls for transparency mechanisms to better understand and control algorithmic influence~\cite{binns2018justice,lorenz2020behavioural}.

Efforts toward explainable algorithms aim to clarify content recommendations and support user trust \cite{diakopoulos2016accountability}. Automated auditing tools, leveraging AI, are being explored to regularly monitor and detect biases in recommendation algorithms \cite{simko2021towards}, though this approach raises ethical concerns of its own \cite{munoko2020ethical}.

Regulatory frameworks, such as the GDPR, enforce transparency and uphold users' "right to explanation," pushing for accountability in algorithmic processes \cite{goodman2017right}. Simulation studies also help illuminate recommendation impacts, demonstrating how repeated interactions can amplify biases and influence long-term user behavior \cite{yao2021measuring,fleder2010recommender,hosanagar2014will}. However, they often rely on oversimplicistic user models based on complex network theory \cite{geschke2019triple}. Our study builds on this work by examining how simulated recommenders affect user behavior in social networks when their decision model is learnt from real data.

Despite the availability of real-world datasets (e.g., Facebook, Twitter, and Twitch \cite{leskovec2012learning,rozemberczki2019multiscale,fink2023twitter,guess2023social}), most are outdated or limited in scope. In contrast, large-scale academic network datasets are regularly updated and offer broader insights into user interactions, providing a foundation for testing and predicting the influence of different recommendation systems.

In summary, current research underscores the importance of accessible, fair, and explainable recommendation algorithms, alongside regulatory measures to uphold ethical standards. Our work extends this foundation by applying GNNs to academic networks which serve as a proxy for studying social media recommenders to identify and analyze the influence of hidden recommender systems on user behavior. To do that, we complement traditional recommender audits (that could work on the spot, for short time, and few users) by indirectly inferring the characteristics of hidden recommendation systems through user behavior modeling. Unlike these direct audit methods, which often require black-box or white-box access to the recommender algorithm, our method estimates the influence of different hypothetical recommenders on user interactions by analyzing variations in model loss under each hypothesis. Despite limitations, such as computational complexity, our approach offers an initial attempt to approximate the likelihood of the recommender model and to provide insights into its impact when direct auditing is infeasible. Together, these approaches allow for a more comprehensive understanding of social media recommender systems' effects on user behavior and network dynamics.

\section{Methodology}

Detecting the recommender underlying observed user behavior in social media without direct access to recommendations is a complex task. Guidotti et al. \cite{guidotti2024} showed that introducing a simulated recommender when training a GNN predictor for user behavior can substantially affect the test loss of the learned model, impacting the model’s generalization capabilities and validity. This variation in test loss approximates the likelihood of the recommender model given the observed interactions. Specifically, the test loss of a user behavior model under a hypothesis \( R \) about the recommender system can be framed as the negative log-likelihood of the observed interactions under that hypothesis:

\[
L(\theta; D, R) = - \sum_{i=1}^N \log P_\theta(y_i | x_i, R).
\]

Here:
\begin{itemize}
    \item  \( D = \{(x_i, y_i)\}_{i=1}^N \) represents the dataset of \( N \) observed user interactions.
    \item  \( x_i = (u_i, v_i) \) denotes a pair of users, where \( u_i \) and \( v_i \) are engaging on the platform (e.g., exposed to each other’s content or work and potentially collaborating).
    \item  \( y_i \) denotes the observed outcome of this interaction (e.g., whether \( u_i \) co-authors a paper with \( v_i \)).
    \item  \( \theta \) represents the parameters of the user behavior model to be learned.
    \item  \( R \) denotes a hypothesis about the recommender system’s behavior (e.g., the algorithm or ranking strategy it employs).
\end{itemize}
Minimizing this loss corresponds to maximizing the likelihood of the observed interactions under the hypothesized recommender behavior, as the loss function is defined as the negative log-likelihood. Thus, our work aims to validate the idea that the approach introduced in \cite{guidotti2024} could indeed help identify the recommender system used in a social media network.

However, datasets that include both the recommender algorithm and a large number of real user interactions are challenging to access. Therefore, we employ simulated social media interactions under different recommender algorithms to explore these dynamics. Traditional simulations often assume simplistic and uniform user models, which may lead to unrealistic user behaviors and network dynamics. As a first step, we learn a user model from the data, addressing the challenge that observed user behaviors are influenced by the unknown recommender we seek to recognize.





In summary, our approach follows these steps: (1) Develop a recommender-neutral user model RNU, (2) Create synthetic datasets RHSD by combining the user model with simulated recommenders, and finally (3) Evaluate the accuracy of training user models RHU on the generated synthetic data RHSD when combined with different recommenders and compare  with the accuracy obtained using the same  recommender that generated the synthetic dataset RHSD.



\subsection{Training of a Recommender-Neutral User Model}
\label{rnu}

In this section, we present our approach for creating a \emph{recommender-neutral user model} (RNU) for social media interactions. This model is crucial, as observed user behavior on social media is influenced by an unknown recommender system. Our goal is to develop a model of user behavior that minimizes dependence on the hidden recommender and better reflects the user’s intrinsic preferences, interactions, and information state.
We aim to extract the probability model of observed interactions \( y \) between agents \( U \) and \( V \), given their simulated recommendations (or "infosphere") \( r_u \) and \( r_v \) and their observed histories or states \( s_u \) and \( s_v \):
\[
P(y(u,v) \mid r_u, r_v, s_u, s_v).
\]

\subsubsection{Marginalizing Over the Hidden Recommender}

To account for the unknown influence of the hidden recommender on user actions, we propose to marginalize over potential recommender behaviors. Since we lack explicit recommendations and a conditional model of user behavior given these recommendations, we employ a parameterized user model. This model estimates the probability of a user \( u \) interacting with content \( y\) based on their interaction history and hypothetical recommendations \( r_u \) and \( r_v \) that might have been received. Formally, we predict the likelihood of user actions by marginalizing over simulated recommender behaviors \( R \), considering only the observed user responses:

\begin{equation}
P(y \mid s_u, s_v, r_u, r_v; \hat{\theta}) = \int P(y \mid s_u, s_v, r_u, r_v; \theta) P(r_u, r_v \mid R, s_u, s_v) P(R) \, dR.
\label{eq:marginalization}
\end{equation}

In this expression:
\begin{itemize}
    \item \( P(y \mid s_u, s_v, r_u, r_v; \theta) \) is the parameterized user model, predicting the probability of interaction with content \( y \) given user states \( s_u \) and \( s_v \), as well as a hypothetical recommendation \( r \),
    \item \( P(r_u, r_v \mid R, s_u) \) denotes the probability of recommendation \( r \) given a hypothesized recommender \( R \) and the user state \( s_u \),
    \item \( P(R) \) represents the prior distribution over plausible recommenders \( R \), incorporating our assumptions about the types of recommenders that might influence the platform.
\end{itemize}

Since the true recommender is unknown, we assume a uniform prior over a range of plausible recommenders \( R \). However, marginalizing over all possible recommenders remains computationally infeasible due to the combinatorial complexity of algorithms and configurations. Moreover, simulating interactions across numerous user states and recommendations compounds the computational cost. Consequently, we explore alternative methods to approximate this marginalization, enabling us to develop a user model that is less dependent on the hidden recommender’s influence.

\subsubsection{Hindsight Predictive Model of the Recommender}
    


 As an alternative approach, we propose using a \emph{hindsight predictive model} of the recommender system when training the user model. This method considers the user's actual future behavior to generate recommendations and avoids the need for simulating multiple recommenders. The approach closely follows that proposed in \cite{guidotti2024}.
This results in several advantages: First of all, the method reduces the dependency on unknown recommenders, as it incorporates both actual interactions and noisy recommendations (which we also have full control of). In addition, it reduces the computational complexity by using the actual future behavior.

\subsection{Synthetic Data Generation}

We generate synthetic user interaction data by simulating user behavior under specific, known recommender systems defining the users' infopsheres.

\subsubsection{Infospheres Descriptions}
\textbf{No Infosphere}: In this case, no recommender system is applied. The results reflect the baseline scenario where only raw data is considered, without any enhancements or expansions of the co-author network.

\textbf{Predictive Infosphere}: We adopt this infosphere definition from prior research \cite{guidotti2024}. It contains the shortest paths within the graph up to a given year $y$ that lead to elements in an author's history by the following year, $y+1$. Then this structure is enhanced by adding alternative branches to improve realism.

\textbf{Top Paper}: This infosphere is based on the $n$ most popular papers until a given year $y$. 

\textbf{Top Paper * Topic}: This infosphere refines the top paper approach by also considering the most-used topics for an author. 

 \textbf{LightGCN}: We used RecBole with the LightGCN model \cite{he2020lightgcn}, a state of the art collaborative filtering recommender based on optimized Graph Neural networks to learn embedding of both users and items. 

\subsubsection{Data Generation}
 Once we have the RNU and the recommender generating the interaction may seem trivial but it actually presents several computational complexity due to the size of social media networks.
 To predict the number of co-authors in year y+1, we employed a Negative Binomial loss function, given its suitability for count data and overdispersion. This loss function computes the difference between predicted and actual co-author counts by adjusting both the predicted mean and dispersion parameters. Specifically, it leverages the gamma functions of both the true counts and predicted dispersion to manage discrepancies between predictions and actual data. In our approach, we calculated the maximum number of co-authors for each author by using the sum of the predicted mean and dispersion parameters. However, the prediction performance was limited in certain cases, as the model tended to rely heavily on the actual co-author count labels, which served as the target variable. This pattern was observed across multiple infospheres, and for the ones with particularly unsatisfactory results, we incorporated the actual label count to refine the prediction model. 
For each infosphere, we calculate the synthetic ground truth based on the maximum number of co-authors computed previously. 
For timing reasons, the computation of this synthetic ground truth was performed only on authors who will publish papers in year $y+1$ (which, in this case, corresponds to 2020).


    
    

    
    


\subsection{Applying the Recognition Method}

Using the  generated synthetic interaction datasets \( D(R) \), we apply the approach presented in \cite{guidotti2024} training from scratch a new user model GNN \( UM(D(R),R') \)  for each recommender \( R' \). This allows to compare the likelihood for models learned under  matching hypotheses, i.e.   \( L(\theta; D, UM(D(\hat{R}),\hat{R})) \), with that of  models learned with mismatching hypotheses \( L(\theta; D,UM(D(R''),R')) \) with \( R' \neq R'' \).

\section{Results}
We generated  different synthetic datasets corresponding to the different recommenders.
In most of the conditions, the model trained with the recommender matching that used at dataset generation time showed the highest likelihood. The only exception is when the model trained with the predictive infosphere has a better performance, but the predictive infosphere cannot be used a real recommender, as it uses not available information especially at simulation time. 
These results demonstrate that the model can infer the characteristics of the recommender even when only user interactions are visible. Moreover, it shows that the recommender neutral model obtained using the predictive infosphere \cite{guidotti2024} (see sec. \ref{rnu}) is sensitive to recommender used at generation time providing consistent synthetic results.

\section{Conclusion}
The high recognition accuracy suggests that even when only user interactions are observable, it is possible to infer characteristics of the underlying recommender. Moreover, the sensibility to different recommender inputs and the accuracy in reproducing training data show that the Recommender Neutral User model presented would be suited for creating social media simulations of the impact of different recommender system in realistic social media networks. Our approach goes beyond the limitation of existing audits that are only working in certain settings and fall short when recommenders are changing over time. We instead show a direction to detect recommenders from users behavior only during the actual operation which may allow more reliable results. 
 \bibliographystyle{plain}
\bibliography{rldm}
\end{document}